\documentclass[pra,twocolumn,superscriptaddress,showpacs]{revtex4-1}
\usepackage{amssymb,amsmath,amsfonts,bm,graphicx}
\usepackage[dvipsnames]{xcolor}
\usepackage{calc,epsfig,epstopdf,color,mciteplus,bm,mathrsfs}
\usepackage{natbib}
\usepackage{times}

\begin{document}

\title{Nonadiabatic holonomic quantum computation with all-resonant control}
% in circuit QED}
%Quantum Electrodynamics}

\author{Zheng-Yuan Xue}\email{zyxue@scnu.edu.cn}   %Corresponding author. E-mail:
\affiliation{Guangdong Provincial Key Laboratory of Quantum Engineering and Quantum Materials, School of Physics\\ and Telecommunication Engineering, South China Normal University, Guangzhou 510006, China}

\author{Jian Zhou} \email{jianzhou8627@163.com}
\affiliation{Department of Electronic Communication Engineering, Anhui Xinhua University, Hefei, 230088, China}
\affiliation{Guangdong Provincial Key Laboratory of Quantum Engineering and Quantum Materials, School of Physics\\ and Telecommunication Engineering, South China Normal University, Guangzhou 510006, China}
\affiliation{National Laboratory of Solid State Microstructure, Nanjing University, Nanjing 210093, China}

\author{Yao-Ming Chu}
\affiliation{School of Physics, Huazhong University of Science and Technology, Wuhan 430074, China}

\author{Yong Hu}
\email{huyong@mail.hust.edu.cn}
\affiliation{School of Physics, Huazhong University of Science and Technology, Wuhan 430074, China}

\date{\today}

\begin{abstract}
The implementation of holonomic quantum computation on superconducting quantum circuits is challenging due to the general requirement of controllable complicated coupling between multilevel systems. Here we solve this problem by proposing a scalable circuit QED lattice with simple realization of a universal set of nonadiabatic holonomic quantum gates. Compared with the existing proposals, we can achieve both the single and two logical qubit gates in an tunable and all-resonant way through a hybrid transmon-transmission-line encoding of the logical qubits in the decoherence-free subspaces. This distinct advantage thus leads to quantum gates with very fast speeds and consequently very high fidelities. Therefore, our scheme paves a promising way towards the practical realization of high-fidelity nonadiabatic holonomic quantum computation.
\end{abstract}


\maketitle

\section{Introduction}

Quantum computation requires a scalable quantum system that can support a universal set of quantum gates.  Being an on-chip implementation, the superconducting quantum circuit (SQC) \cite{decay,JQYou} fulfills the scalable criteria but suffers severely from its environmental fluctuations, which hinder the performance of quantum gates. On the other hand, it is well known that geometric phases and holonomies are largely insensitive to certain local noises as they depend only on the global properties of their cyclical evolution paths. Therefore, holonomic quantum computation (HQC), which exploits the non-Abelian quantum holonomies, has emerged as a promising way towards robust quantum computation \cite{Zanardi1999,Duan2001a,ar,lf,zhangp,ik,jiangl}. As the adiabatic geometric phases demand the adiabatic condition and thus the gate times on the same level of coherence times in typical quantum systems \cite{Wang2001,Zhu2002}, recently, increased theoretical and experimental effort has been attracted by non-adiabatic HQC \cite{Sjoqvist2012, Feng2013, Abdumalikov2013, zu, sac, e7, Wu2005, Xu2012, Liang2014a, Zhang2014d, xu2014,xue,zhouj,wangym}.

However, the realization of nonadiabatic HQC in SQC is far from trivial. Up to now, only single-qubit HQC gates have been experimentally demonstrated \cite{Abdumalikov2013}. Various schemes have been proposed to implement the two-qubit HQC gates, in which one usually needs controllable interaction between addressable multilevel (at least three) systems, and the resulting two-qubit gates are realized in a dispersive manner \cite{ar,Sjoqvist2012}. On the other hand, the anharmonicity of superconducting qubits has been lowered in recent experiments to gain the robustness against $1/f$ noises, limiting the coupling strengths that one can exploit \cite{Abdumalikov2013,mjp}. The required complicated circuit implementation and the relatively slow setup are thus a main obstacle for the realization of holonomic two-qubit gates and consequently the demonstration of universal HQC.

Here we propose a practical scheme for nonadiabatic HQC on a circuit QED lattice.
In our scheme, the logical qubit is encoded in a decoherence-free subspace (DFS) \cite{dfs1} consisting of two transmission-line resonators (TLRs) commonly coupled to a transmon \cite{KochTransmonPRA2007}. The distinct merit of our scheme is that such exotic encoding involves only the lowest two levels of the transmon qubit and can result in universal HQC with all-resonant interactions among the involved elements, leading to fast and high-fidelity universal quantum gates in a very simple setup. In particular, we can obtain a tunable resonant interaction between the transmon and each of the two TLRs through proper ac driving of the transmon, resulting in arbitrary single logical qubit operation. More importantly, for the nontrivial two logical qubit gate, we only need resonant interactions among three TLRs from the two logical qubits, which can be induced by a common grounding superconducting quantum interference device (SQUID) with ac magnetic modulation. Requiring only the current level of technique, our scheme can be immediately tested in experiments and therefore opens up the possibility of realizing universal HQC.

\section{Single qubit gates}

\subsection{The setup and effective Hamiltonian}
\begin{figure}[tbp]
\centering
\includegraphics[width=7cm]{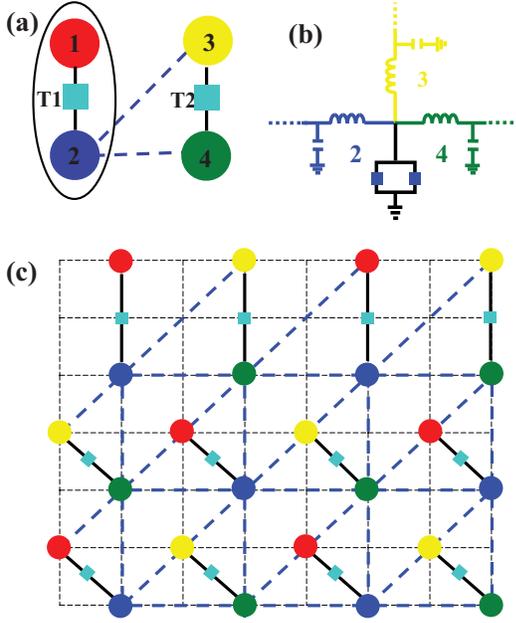}
\caption{(Color online) The proposed setup for our proposal. The circles with different colors denote the TLRs with different frequencies, the squares  denote superconducting transmon qubits, and the black solid and blue dashed bonds indicate that the interactions are for single- and two-qubit gates, respectively. (a) The coupling configuration for a two-qubit gate. The elements in the ellipse denote the encoded first logical qubit. (b) The equivalent circuit for the two-qubit gate consisting of three TLRs coupled by a common grounding SQUID. (c) Scale up to a 2D architecture.} \label{Fig setup}
\end{figure}

The setup we consider is a scalable circuit QED lattice depicted in Fig. \ref{Fig setup}, where the TLRs with different frequencies are denoted by the circles with different colors, the transmon qubits are labeled by the squares, and the capacitive transmon--TLR and inductive TLR--TLR couplings are represented by the solid and dashed bonds, respectively. The logical qubit in our scheme is encoded in the DFS built by two TLRs coupled with a transmon qubit \cite{Steffen2013}, labeled by the ellipse in Fig. \ref{Fig setup}(a) and described by
\begin{eqnarray}\label{Eqn hs}
H_\mathrm{S}= \frac {\omega_q} {2} \sigma^z + \sum_{j=1}^{2}\omega_{c,j} a_j^{\dag} a_j + \sum_{j=1}^{2} \left( g_j a_j\sigma^+ + \mathrm{H.c.}\right),
\end{eqnarray}
where $a_j^{\dag}$ and $a_j$ are the creation and annihilation operators of the $j$th TLR with frequency $\omega_{c,j}$, $\sigma^{z,\pm}$ are the Pauli operators of the transmon qubit with frequency $\omega_q$, and $g_j$ are the real capacitive transmon-TLR coupling strengths.

In the dispersive coupling regime $\Delta_j =\left( \omega_{c,j}-\omega_q \right) \gg g_j$, the resonant transmon--TLR interaction can be induced by biasing the transmon qubit with an ac magnetic flux periodically modulating its transition frequency \cite{Strand2013}. As the transmon qubit is coupled to both of the TLRs, we drive the transmon by a well-controlled two-tone microwave, which results in
\begin{eqnarray}\label{twotone}
\omega_q(t)=\omega_q+ \sum_{j=1}^2 \varepsilon_j \sin (\nu_j t-\phi_j).
\end{eqnarray}
This modulation can effectively tune the sideband of the transmon qubit in resonance with the TLRs. To see this, we move to the rotating frame through $U=U_aU_b$, with
\begin{eqnarray}
U_a&=&\exp\left(-i {\omega_q \over 2} \sigma^z t \right), \notag\\
U_b&=&\exp\sum_{j=1}^{2} \left[i\sigma^z  {\alpha_j \over 2 } \cos (\nu_j t-\phi_j) -i \omega_{c,j} a_j^\dag a_j t \right],
\end{eqnarray}
and $\alpha_j=\varepsilon_j / \nu_j$. The transformed Hamiltonian then reads
\begin{eqnarray}\label{trans}
H_\mathrm{S}^{\prime}(t)&=&\left(g_1 a_1^\dag \sigma^- e^{i \Delta_1 t}
+ g_2 a_2^\dag \sigma^- e^{i \Delta_2 t}\right) \notag\\ &\times& \prod_{j=1}^2 \sum_{m=-\infty}^\infty  i^m J_m(\alpha_j) e^{i m(\nu_j t -\phi_j)} +\mathrm{H.c.},
\end{eqnarray}
with $J_m(\alpha_j)$ being Bessel functions of the first kind. Assuming without
loss of generality $\alpha_2 \simeq 1.4347$ such that $J_0(\alpha_2)=J_1(\alpha_2)=J$ and $g_1=g_2=g/J$, we obtain an effective resonant interaction under the condition $\nu_j=\Delta_j$ as
\begin{eqnarray}  \label{Eqn eff}
H_\mathrm{eff} &=& g[J_1(\alpha_1) a_1^\dag \sigma^-  - J_0(\alpha_1) a_2^\dag \sigma^- e^{i\phi}+ \mathrm{H.c.}],
\end{eqnarray}
with $\phi=\phi_2-\phi_1+\pi$, which indicates the full control over the coupling strength through the design of the a.c. driving of the transmon qubit. Notice that we have neglected the fast-varying terms in deriving Eq. (\ref{Eqn eff}) by the rotating-wave approximation. The omitted term taking the lowest oscillating frequency is
\begin{eqnarray}
\label{Eqn osc}
H_1^{\prime}=g[J_1(\alpha_1) a_2^\dag \sigma^- e^{i\Delta t}
+ J_0(\alpha_1) a_1^\dag \sigma^- e^{-i\Delta t+\phi} + \mathrm{H.c.}]\notag\\
\end{eqnarray}
with $\Delta=\left| \Delta_2 -\Delta_1 \right|$.

\subsection{Universal single qubit gates}

We now move to the construction of the universal set of nonadiabatic holonomic single-qubit quantum gates based on $H_\mathrm{eff}$. Here we consider the DFS
\begin{eqnarray}
S_1&=&\mathrm{span}\{|100\rangle,|001\rangle,|010\rangle\} \notag\\
&\equiv& \mathrm{span} \{|0\rangle_{L},|1\rangle_{L},|E\rangle_{L}\}, \notag
\end{eqnarray}
where $|\alpha \beta \gamma \rangle \equiv |\alpha\rangle_1\otimes|\beta\rangle_q\otimes|
\gamma\rangle_2$ labels the product states of the TLRs and the transmon qubit, the subscript $L$ denotes the states belonging to the logical qubit, and $|E\rangle_{L}$ is an ancillary state. Then, $H_\mathrm{eff}$ in $S_1$ is reduced to
\begin{eqnarray}
\label{Eqn h1}
H_\mathrm{eff}=\lambda_1\left(\sin\frac{\theta}{2}e^{i\phi}|E\rangle_L\langle0|
-\cos\frac{\theta}{2}|E\rangle_L\langle1|\right) + \mathrm{H.c.},
\end{eqnarray}
with $\lambda_1=g\sqrt{J_1^2(\alpha_1)+J_0^2(\alpha_1)}$ and $\tan(\theta/2)
= J_1(\alpha_1)/J_0(\alpha_1)$. Equation (\ref{Eqn h1}) thus establishes in $S_1$ a $\Lambda$-type Hamiltonian from which an arbitrary single-qubit holonomic quantum gate can be resonantly achieved. This can be illustrated in the dressed-state representation where the two lowest states of this three-level system are
\begin{eqnarray}
|d\rangle_{L}&=&\cos\frac{\theta}{2}|0\rangle_{L}+\sin\frac{\theta}{2}
e^{i\phi}|1\rangle_{L}, \notag \\ |b\rangle_{L}&=&\sin\frac{\theta}{2}e^{-i\phi}|0\rangle_{L}-\cos\frac{\theta}{2}|1\rangle_{L}.
\end{eqnarray}
Obviously, the dark state $|d\rangle_{L}$ is decoupled from the other states, while
the bright state $|b\rangle_{L}$ is coupled to the excited state $|E\rangle_{L}$ with effective Rabi frequency $\lambda_1$. When $\lambda_1 \tau_1=\pi$, the dressed states undergo a cyclic evolution in which $| d \rangle_L$ remains invariant and $| b \rangle_L$ evolves to $-| b \rangle_L$. Moreover, as $\langle\psi_{\text{i}}(t)|H_1|\psi_{\text{j}}(t)\rangle=0$ with $|\psi_{\text{i,j}}\rangle\in \{|0\rangle_{L}, |1\rangle_{L}\}$, there is no transition between the two time-dependent states, i.e., the evolution satisfies the parallel-transport condition. Therefore, the evolution operator $U_1=\exp(-i\int_0^{\tau_1} H_1 dt)$ can realize the holonomic operations under the above two conditions. Such evolution can be represented in the subspace $\mathrm{span}\left\{ |0\rangle_{L}, |1\rangle_{L}\right\}$ by
\begin{equation}\label{Eqn u1}
U_1= \left(\begin{array}{ccc}
\cos{\theta}&\sin{\theta}e^{-i\phi}\\
\sin{\theta}e^{i\phi}&-\cos{\theta}
\end{array}\right),
\end{equation}
with $\theta$ and $\phi$ being parameters independently tunable by the ac
driving of the transmon qubit, and thus indicates the implementation of universal single-qubit gates.

\begin{figure}[tbp]
\centering
\includegraphics[width=8cm]{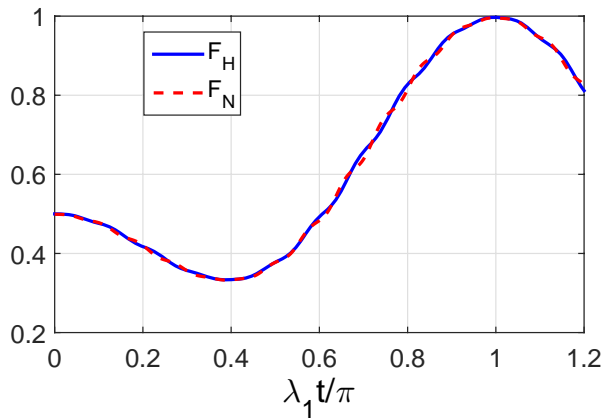}
\caption{(Color online)  Process fidelities $F_H$ and $F_N$ of the Hadamard and the NOT gates, respectively.} \label{Fig single}
\end{figure}

\subsection{Numerical simulation}
The decoherence process in SQC accompanies the described implementation unavoidably, and understanding its effects is crucial for our scheme. The performance of  single-qubit gates is numerically evaluated by the Lindblad master equation,
\begin{eqnarray}
\label{master1}
\dot\rho_1&=& -i[H_\mathrm{eff}+H_1^{\prime}, \rho_1]+\frac \kappa 2
[\mathscr{L}(a_1)+\mathscr{L}(a_2)] \notag \\
&&+\frac {\gamma} {2} \mathscr{L}(\sigma^-)+ \frac {\gamma_{\phi}} {2} \mathscr{L}(\sigma^z),
\end{eqnarray}
where $\rho_1$ is the density matrix of the logic qubit, $\mathscr{L}(\mathcal{A})=2\mathcal{A}\rho_1
\mathcal{A}^\dagger-\mathcal{A}^\dagger \mathcal{A} \rho_1 -\rho_1 \mathcal{A}^\dagger \mathcal{A}$ is the Lindbladian of the operator $\mathcal{A}$, and $\kappa$, $\gamma$, and $\gamma_\varphi$ are the decay rates of the two TLRs, and the relaxation and dephasing rates of the transmon, respectively. For demonstration purposes, we have used in the following a conservative set of experimental parameters. The energy splitting of the transmon and the frequencies of the two TLRs are chosen as $\omega_q/2\pi=6\,\mathrm{GHz}$, $\omega_{c,1}/2\pi=6.5 \mathrm{GHz}$, and $\omega_{c,2}/2\pi=6.75\,\mathrm{GHz}$, which result in $\Delta_1/2\pi=0.5\,\mathrm{GHz}$ and $\Delta_2=1.5\Delta_1$. For the Hadamard gate, we set $\alpha_1 =\varepsilon_1/\Delta_1\simeq 0.7661$ such that $J_0(\alpha_1)=0.8586$ and $J_1(\alpha_1)\simeq \tan(\theta_H/2)J_0(\alpha_1)$. Then we have $\lambda_1/2\pi \simeq 12.73 \,\mathrm{MHz}$ by choosing $g/(2\pi J)=25\,\mathrm{MHz}$. For the NOT gate, our setting is changed to $\alpha_1^\prime\simeq 1.4347$ and thus $\lambda_1^\prime/2\pi \simeq 10.61 \,\mathrm{MHz}$. The decoherence rates of the TLRs and the transmon have all been suppressed to the level $\mathrm{kHz}$ in recent experiments \cite{decay}, and here we set $\kappa=\gamma=\gamma_\phi=2\pi\times10 \mathrm{kHz}$.  Here we choose the Hadamard and the NOT gates as two typical examples which correspond to identical $\varphi=0$ with $\theta_H=\pi/4$ and $\theta_N=\pi/2$ respectively. For an initial state $|\psi_i\rangle=\cos\theta_i|0\rangle_L+\sin\theta_i|1\rangle_L$ of the logical qubit, the Hadamard and the NOT gates should result in the ideal final states $|\psi_H\rangle=[(\cos\theta_i+\sin\theta_i)|0\rangle_L +(\cos\theta_i-\sin\theta_i)|1\rangle_L]/\sqrt{2}$ and $|\psi_N\rangle=\cos\theta_i|1\rangle_L+\sin\theta_i|0\rangle_L$, respectively.  To take systematic errors occurring for any input state into account, we exploit the process fidelity \cite{pf,yin} to characterize the gate performance, which is defined as $F_{H/N}={1\over 2\pi}\int_0^{2\pi} \langle  \psi_{H/N}|\rho_1|\psi_{H/N}\rangle\mathrm{d}\theta_1$. Results shown in Fig. \ref{Fig single} demonstrated that the highest fidelities can be achieved that are $F_H=99.68\%$ and $F_N=99.57\%$, respectively.

It should be emphasized that we have already included in our simulation the fast-varying term in Eq. (\ref{Eqn osc}). Its contribution is at the level of $\pm0.01\%$, which in turn verifies the validity of Eq. (\ref{Eqn eff}). Meanwhile, the influence from the higher-energy levels of the transmon qubit is estimated to be comparable to that of Eq. (\ref{Eqn osc}), and thus can also be safely neglected. In addition, the two-tone modulation in Eq. (\ref{twotone}) is inevitably accompanied by higher harmonics in the qubit frequency modulation. However, we can find out from the derivation of Eq. (\ref{Eqn eff}) that the higher harmonics can only introduce fast-varying terms with the lowest frequency being $2\Delta$, and its contribution is thus even smaller than that of Eq. (\ref{Eqn osc}). Finally, for the counter-rotating terms in Eq. (\ref{Eqn hs}), $g/\{\omega_{\text{q}}, \omega_{\text{c,1}}, \omega_{\text{c,2}} \}<1/240$, the contribution of which is much smaller than that of Eq. (\ref{Eqn osc}). A similar analysis applies also to the latter situation of realizing two-qubit gates.

\section{Nontrivial two-qubit gates}

We next turn to investigate the implementation of the nontrivial two-qubit gates.
For this two logical qubit situation, we exploit the six-dimensional (6D) DFS $S_2$ spanned by
\begin{eqnarray}
&& \{|100100\rangle,|100001\rangle,|001100\rangle,\notag\\
&&|001001\rangle,|101000\rangle,|000101\rangle\} \notag\\
&\equiv & \{|00\rangle_L,|01\rangle_L,|10\rangle_L,|11\rangle_L,|E_1\rangle_L,|E_2\rangle_L\},
\end{eqnarray}
where the former and latter three physical states encode the first and second logical qubits, respectively. To obtain a nontrivial two-qubit gate, we need resonant interaction among three TLRs from two logical qubits, i.e., TLR 2 from logical qubit 1 and TLRs 3 and 4 from logical qubit 2, labeled by the blue dashed bonds in Fig.  \ref{Fig setup}(a). This can be implemented by grounding the involved TLRs at their common end via a SQUID with effective inductance much smaller than those of the TLRs as shown in Fig. \ref{Fig setup}(b). The role of the grounding SQUID is to establish the separated photonic TLR modes and to induce the resonant inter-TLR parametric coupling. The small inductance of the grounding SQUID leads to a low-voltage shortcut for the three TLRs \cite{FelicettiPRL2014} and it is this boundary condition that allows the definition of individual TLR modes on this coupled circuit; see Appendix \ref{appendixa}. In addition, the effective interaction among the three TLRs can be implemented through the dynamic modulation of the Josephson coupling energy of the grounding SQUID via a two-tone ac magnetic flux $\Phi _{\mathrm{ext}}=\Phi_{\mathrm{dc}}+ \sum_{m=1}^2\Phi _m\cos (\omega_m t +\varphi_m)$, where $\Phi_{\mathrm{dc}}$ is the dc bias of the grounding SQUID, and $\Phi_{m}$, $\omega_m$, and $\varphi_m$ are the amplitudes, frequencies, and initial phases of the modulating tones, respectively.

When $\Phi_{m}\ll\Phi_{\mathrm{dc}}$, the resonant parametric coupling between the three TLRs can be obtained in the condition $\omega_1=|\omega_{c,2}-\omega_{c,3}|$ and $\omega_2=|\omega_{c,2}-\omega_{c,4}|$ \cite{FelicettiPRL2014,WangYPChiral2015,Wang2015a}, taking the form of
\begin{equation} \label{twoqubit}
H_{\mathrm{C}}={\eta_1} a_2^\dag a_3 e^{i\varphi_1}+\eta_2 a_2^\dag a_4 e^{i\varphi_2} +\text{H.c.},
\end{equation}
in the rotating frame where $\eta_{m} \propto \Phi _{\mathrm{m}}$ are the hopping strengths tunable by the two modulating tones, see Appendix \ref{appendixb}. Meanwhile, in Appendix {\ref{appendixc}}, we also estimate the fluctuation of the coupling strength induced by the $1/f$ noises, which is shown to be negligibly small. Here we use two tones with different frequencies to mediate the two pairs of TLR-TLR coupling independently. Therefore, the two frequencies should be largely separated from each other to avoid unwanted cross talk. We then choose in our scheme $\omega_1=2\Delta$ and $\omega_2=3\Delta$, which in turn determines the parameters of the second logical qubit as $\omega_{c,3}/2\pi=7.25\,\mathrm{GHz}$, $\omega_{c,4}/2\pi=7.5\,\mathrm{GHz}$, and $\omega_{q,2}=6.75\,\mathrm{GHz}$. Notice that the other single-qubit gate parameters of the logical qubit 2 remain the same as those of the first one with these settings.

In subspace $S_2$, $H_\mathrm{C}$  is reduced to
\begin{eqnarray}\label{h2}
H_\mathrm{C} &=&\lambda_2\left[\sin\frac{\vartheta}{2} e^{i\varphi}
(|E_1\rangle_L\langle00|+|11\rangle_L\langle E_2|) \right. \notag \\
&-& \left. \cos\frac{\vartheta}{2}(|E_1\rangle_L\langle01|+|10\rangle_L\langle E_2|)\right]  +\text{H.c.},
\end{eqnarray}
with $\lambda_2=\sqrt{\eta_{1}^2+\eta_{2}^2}$ being the effective Rabi frequency,
$\tan(\vartheta/2)= \eta_{1}/\eta_{2}$, and $\varphi=\varphi_1-\varphi_2-\pi$. A further inspection shows that $H_\mathrm{C}$ can be decomposed into two commuting parts, i.e., $H_\mathrm{C}=\lambda_{2}(H_a+H_b)$ with
\begin{eqnarray}
H_a&=&\sin\frac{\vartheta}{2}e^{-i\varphi}|E_2\rangle_{L}\langle11|
-\cos\frac{\vartheta}{2}|E_2\rangle_{L}\langle10|+\mathrm{H.c.}, \notag\\
H_b &=&\sin\frac{\vartheta}{2}e^{i\varphi}|E_1\rangle_{L}\langle00|
-\cos\frac{\vartheta}{2}|E_1\rangle_{L}\langle01|+\mathrm{H.c.}.
\end{eqnarray}
It is noticed that both $H_a$ and $H_b$ take the similar form of $H_\mathrm{eff}$
in Eq. (\ref{Eqn h1}), acting nontrivially on their individual computational subspaces $\mathrm{span}\{|10\rangle_{L},|11\rangle_{L}\}$ and $\mathrm{span}\{|00\rangle_{L},|01\rangle_{L}\}$, respectively. Therefore, the holonomic two-qubit logical gate,
\begin{eqnarray}
U_2(\vartheta, \varphi)=\left(
  \begin{array}{cccc}
    \cos\vartheta & \sin\vartheta e^{-i\varphi} & 0 & 0 \\
    \sin\vartheta e^{i\varphi} & -\cos\vartheta & 0 & 0 \\
    0 & 0 & -\cos\vartheta & \sin\vartheta e^{-i\varphi} \\
    0 & 0 & \sin\vartheta e^{i\varphi} & \cos\vartheta \\
  \end{array}
\right),\notag\\
\end{eqnarray}
can be obtained with parameters $\vartheta$ and $\varphi$ tunable by the external two-tone modulation.

\begin{figure}[tbp]
\centering
\includegraphics[width=6.5cm]{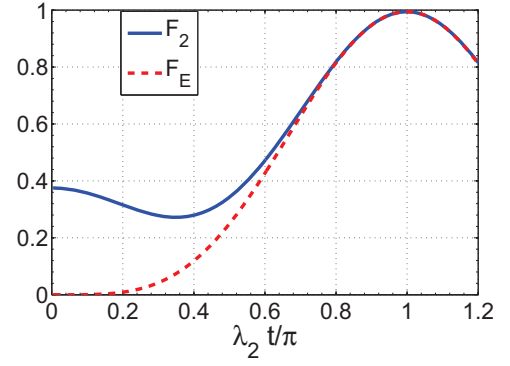}
\caption{(Color online) Performance of the two-qubit gate $U_2(\pi/4, 0)$.  Process fidelity $F_2$ for  $U_2(\pi/4, 0)$ (blue solid line) and state fidelity $F_E$ for $|\psi_2'\rangle=(|0\rangle_1+|1\rangle_1)|1\rangle_2/\sqrt{2}$ (red dashed line).}\label{twobit}
\end{figure}

%\subsection{Numerical simulation}
We further verify the performance of the two-qubit gates by taking $U_2(\pi/4,0)$ as an example. Here we set  $\eta_1/2\pi=4.14\,\mathrm{MHz}$ and $\eta_2/2\pi=10\,\mathrm{MHz}$ such that $\eta_1/\eta_{2} \simeq \tan(\pi/8)$ and $\lambda_2/2\pi = 14.14\,\mathrm{MHz}$. %As shown in Fig. \ref{twobit}(a),
We calculate the state populations and fidelity for an initial state $|01\rangle_L$ and obtain the fidelity with its highest being $F_T=99.42\%$. Moreover, similar to the single-qubit case, we calculate the process fidelity $F_2=(4\pi^2)^{-1}\int_0^{2\pi} \int_0^{2\pi} \langle \psi_f|\rho_2|\psi_f\rangle d\vartheta_1d\vartheta_2$ for an initial state $|\psi_2\rangle=(\cos\vartheta_1|0\rangle_1+\sin\vartheta_1|1\rangle_1)
(\cos\vartheta_2|0\rangle_2+\sin\vartheta_2|1\rangle_2)$, with obtained results
approaching $99.42\%$, as shown in Fig. \ref{twobit}. Here $\rho_2$ is the numerically simulated density matrix of the two logical qubit system and $|\psi_f\rangle=U_2(\pi/4,0)|\psi_2\rangle$. The obtained fidelity is comparable to that of the single-qubit operations, which is in sharp contrast to the existing implementations and can be interpreted in an intuitive way: Since all the interactions exploited in our scheme are resonant, the effective two-qubit coupling strength $\lambda_2$ is of the same order of the single-qubit $\lambda_1$, leading to high-fidelity two-qubit quantum gates. This all-resonant feature is distinct from the previous schemes where two-qubit gates are usually induced by dispersive interactions which leads to longer time durations and lower gate fidelities. In addition, as the $U_2(\pi/4,0)$ gate is an entangling gate, i.e., the ideal final state is a maximum entangled two-qubit state when the initial state is $|\psi_2'\rangle=(|0\rangle_1+|1\rangle_1)|1\rangle_2/\sqrt{2}$, we plot the state fidelity of $|\psi_2'\rangle$ in  Fig. \ref{twobit} to characterize the entangling nature of this gate, where we can obtain $F_E=99.42\%$ and verify that the final state is a nearly maximum entangled state.

\section{Discussion}

%\subsection{Scalability}
Our scheme of single-qubit and two-qubits gates can be easily scaled up to facilitate the scalability criteria of quantum computing. As shown in Fig. \ref{Fig setup}(c), we can form a 2D array of the logic qubits by placing the TLRs in an interlaced form. Such scaled array involves several SQUIDs which should be controlled by external magnetic flux bias (both the SQUIDs of the transmon qubit and the grounding SQUIDs). Meanwhile, this requirement does not place a hindrance on the future experimental realization of the proposed scheme. In the past decade, the individual flux control has already been achieved in coupled superconducting flux qubits \cite{Plant,Ploeg}, where several coils have been applied to manipulate the dc and ac magnetic fluxes threaded in neighboring flux qubits with both the loop sizes of the qubits and the distances between the qubits being at the range of micrometers. On the other hand, the spacing between the building elements in our proposal is at the same length scale of the TLRs (millimeters; see the table in Appendix \ref{appendixa}), which is by several orders larger than the case of the coupled flux qubits. From this point of view, the requirement of individual flux addressing in our scheme is weaker than those of the reported experiments because the larger distance between the SQUID loops indicates smaller cross talk and easier fabrication of the biasing coils. When the scaled-up lattice is taken into consideration, the requirement of controlling many SQUIDs individually leads to more complicated coil setup than that of the few-qubit case. However, the very large spacing between the grounding SQUIDs still offers enough room for the circuit design. One potential solution is that we may add an additional layer of antenna on top of the sample that contains the array of the TLRs. Here we should notice that increasing research interest has recently been attracted by the design of scalable architecture that combines various quantum elements into a complex device without compromising their performance, and a multilayer microwave integrated quantum circuit platform has already been developed to couple a large number of circuit components through controllable channels while suppressing any other interactions \cite{Multilayer,m2}. In addition, the parametric coupling method exploited here has been investigated in various recent experiments, where the Hong-Ou-Mandel interference and the synthetic gauge field for the microwave photons in the TLRs have been observed \cite{NISTHongOuMandelPRL2012,Roushan}. These experimental advances thus partially verify the feasibility of our scheme.

\section{Conclusion}
In summary, we have proposed to implement universal HQC in DFS on a circuit QED lattice. Through the control of the amplitudes and relative phases of the ac
magnetic modulating flux, arbitrary single- and two-qubit gates can be resonantly realized. Such speedup thus pushes the gate fidelities in the presence of decoherence to unprecedented high level. Therefore, our scheme paves a promising way towards the realization of high-fidelity HQC in superconducting circuits.

\acknowledgments We thank Dr. Shi-Lei Su, Professor Yan-Kui Bai and Professor Ming Yang for helpful discussion. This work is supported by the NFRPC (Grant No. 2013CB921804), the NKRDPC (Grant No. 2016YFA0301803), the NSFC (Grant No. 11374117), the PCSIRT (Grant No. IRT1243), the Education Department of Anhui Province (Grant No. KJ2015A299), and NJU (Grant No. M28015).

\appendix

\section{Eigenmodes of the building block}
\label{appendixa}

We begin with the case that only a dc flux bias $\Phi_\mathrm{dc}$ is added.   The Lagrangian of the building block can be written as
\begin{align}
\mathcal{L}&= \sum_{\alpha} \int_{0}^{L_{\alpha}}\, \mathrm{d}x \, \frac{1}{2} \left[c\left(\frac{\partial \phi_{\alpha}(x,t)}{\partial t}\right)^2 -\frac{1}{l}\left(\frac{\partial \phi_{\alpha}(x,t)}{\partial x}\right)^2\right]  \notag\\
&+\frac{1}{2}C_{\mathrm{J}}\dot{\phi}_{\mathrm{J}}^2
+E_{\mathrm{J}}\cos \left(\frac{ \phi_{\mathrm {J}} }{\phi_{\mathrm{0}}}\right)\\
&\approx \sum_{\alpha} \int_{0}^{L_{\alpha}}\, \mathrm{d}x \, \frac{1}{2}\left[c\left(\frac{\partial \phi_{\alpha}(x,t)}{\partial t}\right)^2
-\frac{1}{l}\left(\frac{\partial \phi_{\alpha}(x,t)}{\partial x}\right)^2\right]  \notag\\
&+\frac{1}{2}C_{\mathrm{J}} \dot{\phi}_{\mathrm{J}}^2-\frac{1}{2L_{\mathrm{J}}}\phi_{\mathrm{J}}^2
\label{eq:Lagrangian2}
\end{align}
with $c$ and $l$ being the capacitance and inductance per unit length of the TLRs, $\alpha=2,3,4$ the label of the three TLRs, $L_{\alpha}$ the length of the $\alpha$th TLR,  $C_{\mathrm{J}}$ the capacitance of the SQUID with the effective Josephson energy being $E_{\mathrm{J}}=E_{\mathrm{J0}} \cos( \pi \Phi_{\mathrm{ext}} / \Phi_{\mathrm{0}})$ with $E_{\mathrm{J0}}$ being its maximal Josephson energy, $\Phi_{\mathrm{ext}}$ the external flux bias, and $\Phi_{\mathrm{0}}=h/2e$ the flux quantum. $\phi_{\mathrm{0}}=\Phi_{0}/2\pi$ is the reduced flux quantum, $L_{\mathrm{J}}=\phi_{\mathrm{0}}^2/E_{\mathrm{J}}$ is the effective inductance of the SQUID, $V_{\alpha}(x,t)$ is the voltage distribution on the TLR $\alpha$, $\phi_{\alpha}(x,t)=\int_{-\infty}^{t} \mathrm{d}t' \,V_{\alpha}(x,t')$ is the corresponding node flux distribution, $V_{\mathrm{J}}(t)$ is the voltage across the grounding SQUID, and $\phi_{\mathrm{J}}(t)=\int_{-\infty}^{t} \mathrm{d}t' \, V_{\mathrm{J}}(t')$. In deriving Eq.~(\ref{eq:Lagrangian2}), we have linearized the grounding SQUID as $E_{\mathrm{J}}\cos(\phi_{\mathrm{J}}/\phi_{\mathrm{0}})\approx -\phi_\mathrm{J}^2/2L_\mathrm{J}$, which is consistent with the described shortcut boundary condition.

\begin{figure}[b]
\centering
\includegraphics[width=8cm]{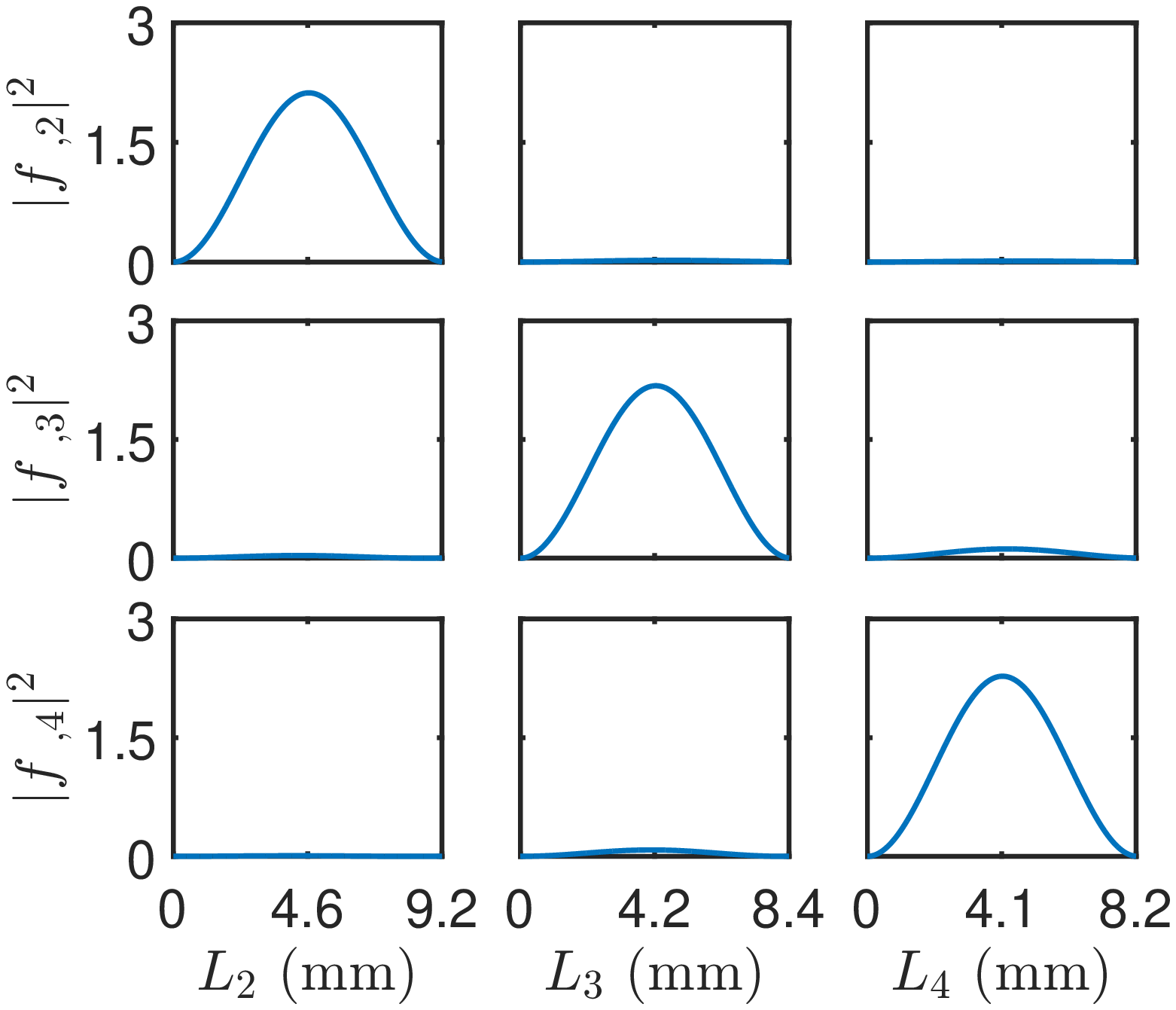}
\caption{(Color Online) Normalized node flux distributions of the three eigenmodes of the coupled TLRs. $L_\alpha$ and $|f_{\alpha,n}|^2$ are in units of $\mathrm{mm}$ and $10^2 \mathrm{m^{-1}}$, respectively.}
\label{Fig Eigenmode}
\end{figure}

The equation of motion of $\phi_\alpha$ has the wave equation form,
\begin{align}
\frac{\partial ^{2}\phi _{\alpha }}{\partial x^{2}}-\frac{1}{v^{2}}\frac{\partial^{2}\phi _{\alpha }}{\partial t^{2}}=0, \label{eq:wavequation}
\end{align}
with $v=1/\sqrt{cl}$, and the boundary conditions are obtained, from Kirchhoff's law, as
\begin{eqnarray}
\phi_{\alpha}(x=0)=0,\,\phi_{\alpha}(x=L_{\alpha})=\phi_\mathrm{J}, \label{eq:boundaryconditon01}\\
-\frac{1}{l}\sum _{\alpha }\frac{\partial \phi_{\alpha} }{\partial x} |_{x=L_{\alpha}} =\frac{\phi _{\mathrm{J}}}{L_{\mathrm{J}}}+C_{\mathrm{J}}\ddot{\phi}_{\mathrm{J}}. \label{eq:boundaryconditon02}
\end{eqnarray}
The variable separation ansatz $\phi_{\alpha}(x,t)=\sum_{m}f_{\alpha,m}(x)g_m(t)$ is then exploited, with $m=2,3,4$ being the index of the eigenmodes.  We can obtain $f_{\alpha,m}(x)=C_{\alpha,m}\sin(k_m x)$ from Eq.~(\ref{eq:boundaryconditon01}), and by inserting $f_{\alpha,m}(x)$ into Eq.~(\ref{eq:boundaryconditon02}), we get
\begin{align}\label{eq:transcendental}
& \sum _{\beta}C_{\beta,m }L_{\mathrm{J}}k_m\cos \left( k_mL_{\beta }\right) \notag \\
+{} &\left( l-\frac{C_{\mathrm{J}}L_{\mathrm{J}}}{c}k_m^{2}\right) C_{\alpha,m}\sin (k_mL_{\alpha})=0,
\end{align}
from which $f_{\alpha,m}(x)$ can be solved up to a constant and the typical solution is plotted in Fig.~\ref{Fig Eigenmode}, which  indicates the one-to-one correspondence between the TLRs and the eigenmodes. In the numerical solution, the parameters are chosen from recent experiments in circuit QED, as listed in Table \ref{Tab para}.

\begin{table}[tb]
\centering
\caption{Representative parameters of the proposed circuit which are selected based on recent reported experiments of parametric processes in circuit QED.}
\label{Tab para}
\begin{tabular}{p{0.23\textwidth}p{0.24\textwidth}}
  \hline  \hline
  \hspace{.1cm} TLRs parameters &  \\
  \hline
   \hspace{.1cm} unit inductance/capacitance & $l=4.1\times 10^{-7}\,
   \mathrm{H}\mathrm{m}^{-1}$, $c=1.6\times 10^{-10}\,\mathrm{F}\mathrm{m}^{-1}$ \cite{NISTParametricConversionNP2011,NISTHongOuMandelPRL2012,NISTParametricCouplingPRL2014}\\
   \hspace{.1cm} lengths of the TLRs &  $L_2=9.16\,\mathrm{mm}$, $L_3=8.46\,\mathrm{mm}$, $L_4=8.2\,\mathrm{mm}$ \cite{DCEexperimentNature2011,NISTParametricConversionNP2011,NISTHongOuMandelPRL2012}\\
  \hline
   \hspace{.1cm} SQUID &\\
  \hline
   \hspace{.1cm} maximal critical current & $I_{\mathrm{J0}}=29.5\,\mu \mathrm{A}$ \cite{DCEexperimentNature2011,NISTParametricConversionNP2011,
  YuYangScience2002,MartinisPhaseQubitPRL2002}\\
   \hspace{.1cm} d.c. flux bias point & $\Phi_\mathrm{dc}=0.33\Phi_0$ \cite{NISTParametricConversionNP2011,NISTHongOuMandelPRL2012}\\
   \hspace{.1cm} effective critical current & $I_{\mathrm{J}}=15\,\mu \mathrm{A}$\\
   \hspace{.1cm} junction capacitance & $C_\mathrm{J}=0.5\,\mathrm{pF}$ \cite{YuYangScience2002,MartinisPhaseQubitPRL2002}\\
   \hspace{.1cm} a.c. modulation amplitudes & $\Phi_\mathrm{23}=0.5\%\Phi_0$, $\Phi_\mathrm{24}=1.5\%\Phi_0$ \cite{NISTParametricConversionNP2011}\\
  \hline
   \hspace{.1cm} Eigenmodes \& coupling & \\
  \hline
   \hspace{.1cm} eigenfrequencies & $\omega_{c2}/2\pi=6.75\,\mathrm{GHz}$, $\omega_{c3}/2\pi=7.25\,\mathrm{GHz}$, $\omega_{c4}/2\pi=7.5\,\mathrm{GHz}$\\
  \hspace{.1cm}  uniform decay rate & $\kappa/2\pi=10\,\mathrm{kHz}$ \cite{DCEexperimentNature2011,NISTParametricConversionNP2011,YuYangScience2002, MartinisPhaseQubitPRL2002,NISTCoherentStateAPL2015}\\
   \hspace{.1cm} hopping strengths & $\eta_{1}/2\pi=4.14\,\mathrm{MHz}$, $\eta_{2}/2\pi=10\,\mathrm{MHz}$\\
  \hline  \hline
\end{tabular}
\end{table}

The quantization of the eigenmodes is then straightforward.  Through the definition of the creation and annihilation operators,
\begin{align}
a_{m}^{\dag } =\sqrt{\frac{\omega_{c,m} c}{2\hbar }}g_{m}-i\sqrt{\frac{1}{%
2\hbar \omega_{c,m} c}}\pi _{m},\notag \\
a_{m} =\sqrt{\frac{\omega_{c,m} c}{2\hbar }}g_{m}+i\sqrt{\frac{1}{2\hbar \omega_{c,m} c }}\pi _{m},
\end{align}
$\mathcal{H}_0$ can finally be written as
\begin{align}
\mathcal{H}_{\mathrm{0}}=\sum_{m}\hbar \omega _{c,m}
\left(a_{m}^{\dag }a_{m}+\frac{1}{2}\right).
\label{eq:eigenhamiltonian}
\end{align}
In particular, $\phi_\mathrm{J}$ can be written as
\begin{equation}
\label{eq:phiJ1}
\phi_\mathrm{J}=\sum_{m} \phi^{m}(a_m+a_m^\dagger),
\end{equation}
with $\phi^{m}=f_{\alpha,m}(x=L_\alpha)\sqrt{\hbar/2\omega_{c,m} c}$ being the rms node flux fluctuation of the $m$th mode across the grounding SQUID. With the parameters in Table \ref{Tab para}, we estimate that
\begin{align}
(\phi^2,\phi^3,\phi^4)/\phi_0
= (3.3,3.4,2.3)\times 10^{-3}.
\end{align}
Such small fluctuation of $\phi_\mathrm{J}$ indicates that the eigenmodes can be regarded as the individual $\lambda/2$ modes of the TLRs, which are slightly mixed by the grounding SQUID with small but finite inductance; see, also, Fig.~\ref{Fig Eigenmode}. Meanwhile, the small $\phi^{m}$ also validate the linearization of  the grounding SQUID.

\section{Parametric coupling between the eigenmodes}
\label{appendixb}

The parametric coupling between the three eigenmodes originates from the dependence of $E_{\mathrm{J}}$ on $\Phi_{\mathrm{ext}}$. Assuming that a small ac fraction $\Phi_{\mathrm{ac}}(t)$ has been added to $\Phi_{\mathrm{ext}}$,
\begin{align}
E_{\mathrm{J}}&=E_{\mathrm{J0}}\cos \left[\frac{1}{2\phi _{0}}\left(\Phi_{\mathrm{dc}}+\Phi_{\mathrm{ac}}(t)\right)\right]   \notag\\
&\approx E_{\mathrm{J0}}\cos \left(\frac{\Phi_{\mathrm{dc}}}{2\phi _{0}}\right)-\frac{E_{\mathrm{J0}}\Phi_{\mathrm{ac}}(t)}{2\phi _{0}}\sin \left(\frac{\Phi_{\mathrm{dc}}}{2\phi _{0}}\right),
\label{Eqn EJosillation2}
\end{align}
where $\left|\Phi_{\mathrm{ac}}(t)\right| \ll \left|\Phi_{\mathrm{dc}} \right|$.
As stated in the main text, $\Phi_{\mathrm{ac}}(t)$ is composed of two tones,
\begin{align}
\label{Eqn twotone}
\Phi_{\mathrm{ac}}(t)&=\Phi_2\cos ( 3 \Delta t + \varphi_2 )+\Phi_1\cos ( 2 \Delta t + \varphi_1 )
\end{align}
where the $3\Delta$ and $2\Delta$ tones are exploited to induce the $2 \Leftrightarrow 4$  and $2 \Leftrightarrow 3$ hopping, respectively. By representing $\phi_{\mathrm{J}}$ as the form shown in Eq.~(\ref{eq:phiJ1}), we obtain the ac coupling from the second term of Eq.~ (\ref{Eqn EJosillation2}) as
\begin{equation}
\mathcal{H}_{\mathrm{ac}}=\frac{E_{\mathrm{J0}}\Phi_{\mathrm{ac}}(t)}{4\phi _{0}^3}\sin \left(\frac{\Phi_{\mathrm{dc}}}{2\phi _{0}}\right)
\left[\sum _{m} \phi^m \left(a_m+a_m^{\dagger}\right)\right]^2,
\label{eq:HDCcoupling2}
\end{equation}
In the rotating frame with respect to $\mathcal{H}_{\mathrm{0}}$, the induced parametric coupling among the TLRs can be obtained as  \begin{equation} \label{twoqubit}
H_{\mathrm{C}}={\eta_1} a_2^\dag a_3 e^{i\varphi_1}+\eta_2 a_2^\dag a_4 e^{i\varphi_2} +\text{H.c.},
\end{equation}
and the fast-oscillating terms in $e^{it\mathcal{H}_{\mathrm{0}}} \mathcal{H}_{\mathrm{ac}} e^{-it\mathcal{H}_{\mathrm{0}}}$ are omitted due to the rotating-wave approximation. When $\left[ \Phi_1, \Phi_2\right]=\Phi_0\left[ 0.5\%, 1.5\% \right]$, the coupling strengths $\eta_1/2\pi=4.14\,\mathrm{MHz}$ and $\eta_2/2\pi=10\,\mathrm{MHz}$ can be induced \cite{NISTParametricConversionNP2011,NISTHongOuMandelPRL2012,
NISTParametricCouplingPRL2014,NISTCoherentStateAPL2015}.

We should also note that the modulating frequency of $\Phi_{\mathrm{ac}}(t)$ must be lower than the plasma frequency of the grounding SQUID $\omega_{\mathrm{p}}=\sqrt{8E_{\mathrm{C}} E_\mathrm{J}}$ \cite{KochTransmonPRA2007}, otherwise the internal degrees of freedom of the SQUID will be activated and complex quasiparticle excitations will emerge. This requirement is guaranteed by the very small inductance of the grounding SQUID. With the parameters selected we have the estimation $\omega_{\mathrm{p}} \approx 2\pi \times 85 \mathrm{GHz}=340\Delta$, leading to the effective suppression of the grounding SQUID excitation.

\section{The influence of $1/f$ noises} \label{appendixc}
We further estimate the fluctuation of the coupling strength induced by the $1/f$ noises, which is ubiquitous in SQC and whose influence exceeds that of the thermodynamic noise \cite{FlickerRMP2014}. The $1/f$ noise originates mainly from the fluctuations of three degrees of freedom, namely the charge, the flux, and the critical current. First, the proposed circuit is insensitive to the charge noise as it consists of only linear TLRs, grounding SQUIDs with very small anharmonicity and the charge-insensitive transmon qubits \cite{KochTransmonPRA2007}. Second, for the flux-type $1/f$ noise, various previous measurements has shown its strength falls in $\mathcal{A}_\Phi/\Phi_0 \in  \left[10^{-6},10^{-5}\right]$ and does not vary greatly with the loop size, inductor value, or temperature \cite{FluxqubitFlickerPRL2006,MartinisFlickerPRL2007,FluxQubit1fGeometryPRB2009}. Therefore, the strength of $\delta \Phi$ can be estimated as $\delta \Phi/\Phi_0 \in [10^{-5},10^{-4}]$, which is by two orders of magnitude smaller than both the dc and ac bias of the two-tone modulation. The existence of $\delta \Phi$ shifts $\Phi_\mathrm{dc}$ in a quasistatic way, and based on the experimental parameters  listed in the table of Appendix \ref{appendixa}, we can evaluate that  $\delta \Phi$ causes negligible
\begin{eqnarray}
\delta \omega_m &\in 2\pi\times [10^{-3},10^{-2}] \text{ MHz} < 10^{-3} \eta_2, \notag\\
\delta \eta_{1,2} &\in 2\pi\times [10^{-4},10^{-3}] \text{ MHz} < 10^{-4} \eta_2.
\end{eqnarray}
In addition, experiments have shown that the critical current noise has $\mathcal{A}_{I_\mathrm{J0}} \approx 10^{-6} I_{\mathrm{J0}}$ for a junction
at temperature $4\,\mathrm{K}$ \cite{JohnClarkICPRB2004,MartinisFlickerPRL2007}. The parameter $\mathcal{A}_{I_\mathrm{J0}}/I_\mathrm{J0}$ is proportional to the temperature down to at least $100\,\mathrm{mK}$. Therefore, we can set $\mathcal{A}_{I_\mathrm{J0}}/I_\mathrm{J0} \in [10^{-7},10^{-6}]$. The influence of the critical current noise can also be estimated and we find that $\delta I_{\mathrm{J0}}$ causes
\begin{eqnarray}
\delta \omega_m &\in 2\pi\times [10^{-4},10^{-3}] \text{ MHz} < 10^{-4} \eta_2, \notag\\
\delta \eta_{1,2} &\in 2\pi\times [10^{-5},10^{-4}] \text{ MHz} < 10^{-5} \eta_2,
\end{eqnarray}
which are even smaller than the flux noise effects.

\end{document}